\documentclass[12pt]{article}
\usepackage{amssymb,amsfonts}

\usepackage{verbatim}
\usepackage{cite}
\usepackage{amsmath}
\usepackage{lscape}
\usepackage{epsfig}

\def\trace{\mathop{\rm Tr}\nolimits}

\def\rmd{{\rm d}}

\providecommand{\openone}{\leavevmode\hbox{\small1\kern-3.8pt\normalsize1}}

\csname @addtoreset\endcsname{equation}{section}

\providecommand{\href}[2]{#2}

\newcommand{\SO}{\mathop{\rm SO}}
\newcommand{\U}{\mathop{\rm {}U}}

\newcommand{\mf}{\|\varphi\|}
\newcommand{\beqn}{\begin{eqnarray}}
\newcommand{\eeqn}{\end{eqnarray}}

\newcommand{\ft}[2]{{\textstyle\frac{#1}{#2}}}

\def\trace{\mathop{\rm Tr}\nolimits}

\def\rmd{{\rm d}}

\newcommand{\be}{\begin{equation}}
\newcommand{\ee}{\end{equation}}
\newcommand{\bea}{\begin{eqnarray}}
\newcommand{\eea}{\end{eqnarray}}
\newcommand{\bref}[1]{(\ref{#1})}

\def\del3{\delta^{(3)}}

\def\otaula{\begin{tabular}}
\def\ctaula{\end{tabular}}

\def\tI{\tilde{I}}
\def\tJ{\tilde{J}}
\def\tK{\tilde{K}}
\def\calM{\mathcal{M}}
\def\modph{\|\varphi\|}

\def\CC{{\mathchoice
{\rm C\mkern-8mu\vrule height1.45ex depth-.05ex
width.05em\mkern9mu\kern-.05em}
{\rm C\mkern-8mu\vrule height1.45ex depth-.05ex
width.05em\mkern9mu\kern-.05em}
{\rm C\mkern-8mu\vrule height1ex depth-.07ex
width.035em\mkern9mu\kern-.035em}
{\rm C\mkern-8mu\vrule height.65ex depth-.1ex
width.025em\mkern8mu\kern-.025em}}}

\def\RR{{\rm I\kern-1.6pt {\rm R}}}

\def\ZZ{{\rm Z}\kern-3.8pt {\rm Z} \kern2pt}

\def\tn{\tilde{n}}
\def\tx{\tilde{x}}
\def\ty{\tilde{y}}

\begin{document}

\begin{titlepage}
\begin{flushright}
KUL-TF-05/03\\
hep-th/0502202
\end{flushright}
\vspace{0.5cm}
\begin{center}
\baselineskip=16pt {\LARGE Stable de Sitter vacua in N=2, D=5
supergravity \\}
\vfill
{\Large Bert Cosemans and Geert Smet  $^\dagger$ 
  } \\
\vfill
{\small Instituut voor Theoretische Fysica, Katholieke Universiteit Leuven,\\
       Celestijnenlaan 200D B-3001 Leuven, Belgium.
 }
\end{center}
\vfill
\begin{center}
{\bf Abstract}
\end{center}
{\small We find $5D$ gauged supergravity theories exhibiting stable de Sitter vacua.  These are the first examples of stable de Sitter vacua in higher-dimensional ($D>4$) supergravity.
Non-compact gaugings with tensor multiplets and R-symmetry gauging seem to be the essential ingredients in these models. They are however not sufficient to guarantee stable de Sitter vacua, as we show by investigating several other models. The qualitative behaviour of the potential also seems to depend crucially on the geometry of the scalar manifold.
 }\vspace{2mm} \vfill
 \hrule width 3.cm
 {\footnotesize \noindent
$^\dagger$  \{bert.cosemans, geert.smet\}@fys.kuleuven.ac.be }
\end{titlepage}
\newpage
\section{Introduction}
Cosmological observations \cite{Cosmo}, suggesting that our
current universe has a small positive cosmological constant, have
lead to a vigorous search for de Sitter vacua in string theory
(see e.g. \cite{dsString}) and more modestly in supergravity (see
e.g \cite{dsSugra,stabledSa,stabledSb,dsN4a,dsN4b,Kallosh}).  Up to this day, the
only examples of stable de Sitter vacua in extended ($N \geq 2$)
supergravity were found by Fr\' e et al in \cite{stabledSa} in the
context of $N=2$ $D=4$ gauged supergravity. Some very specific
elements of 4D supergravity were used to construct these examples,
some of which have no clear string theory origin (like for
instance the de Roo-Wagemans angles). The embedding of their
models in $N=4$ $D=4$ gauged supergravity, and all semi-simple
gaugings of $N=4$ $D=4$ supergravity coupled to six vector multiplets were discussed in
\cite{dsN4a,dsN4b}, but all the de Sitter vacua turned out to be
unstable.  In view of these problems, we thought it might be
instructive to look for de Sitter vacua in higher-dimensional
gauged supergravity theories, in order to find out what general
ingredients are necessary to guarantee the existence of stable de Sitter
vacua.

In this paper we will focus on $5D$ $N=2$ gauged supergravity for
several reasons.  First of all, it is very similar to $4D$ $N=2$ in
certain respects. They both allow an arbitrary number of vector-
and hypermultiplets, and there exist beautiful relations between
their respective scalar manifolds. On the other hand, tensor
multiplets seem to be somewhat easier to introduce
in $5D$, and there are no duality symmetries in $5D$, which makes
the $5D$ theory simpler \footnote{For interesting recent progress on the
coupling of scalar-tensor multiplets to $4D$ $N=2$ supergravity,
see \cite{scaltens}. The coupling of a vector-tensor multiplet to supergravity was
done in \cite{VecTen}.}.

Besides this, there are some very good other reasons to study $5D$
gauged supergravity. An important motivation comes from the
holographic principle, of which AdS/CFT \cite{AdsCft} is a
particularly nice realization.  The best understood example is the
famous correspondence between Type $IIB$ string theory on $AdS_5
\times S^5$ and $N=4$ Super Yang Mills theory.  A lot can be
learned just by looking at the $5D$ $N=8$ SO(6) gauged
supergravity (which is assumed to be a consistent truncation of
$IIB$ supergravity on $AdS_5 \times S^5$).  For instance, Anti-de Sitter
critical points of the potential imply, under suitable conditions,
non-trivial conformal fixed points for the Yang Mills theory, and
supergravity kink solutions that interpolate between Anti-de Sitter vacua
correspond to renormalization group flows in the dual YM theory
(see \cite{RGflow} for a short review).  There has been a lot of
speculation that similarly, de Sitter (quantum) gravity might be
dual to some (still unknown) Euclidean conformal field theory (see
e.g \cite{dsCft}).  However, the correspondence is a lot less
clear than in the Anti-de Sitter case (for some recent reviews see
\cite{dsCft2}).  Studying gauged supergravities (with probably
non-compact gauge groups) that have stable de Sitter vacua might give us
some more clues about a possible dS/CFT correspondence.

Finally, from a more phenomenological point of view, we should
note that various authors have suggested that the universe may
have undergone a phase where it was  effectively 5-dimensional,
see e.g. \cite{5Dpheno,options} and references therein, giving
another reason to understand the vacuum structure of $5D$ ($N=2$)
gauged supergravity theories.

Our paper is organized as follows. In section 2 we repeat some
elements of 5D gauged supergravity coupled to tensor and vector
multiplets.  We then study the potential corresponding to
R-symmetry gauging in more detail in section 3, where we prove
that $U(1)_R$ gauging does not give rise to stable de Sitter
vacua. In section 4 we present the first examples of stable de Sitter
vacua in 5D gauged supergravity. Tensors charged under a
non-compact group and R-symmetry gauging seem to be crucial. The
following section contains more examples of $5D$ $N=2$ gauged
supergravities with charged tensors and R-symmetry gauging.
Unfortunately, these do not lead to stable de Sitter vacua.  In section 6
we show that if we replace R-symmetry gauging by a specific $U(1)$
gauging of the universal hypermultiplet, we can also get stable de Sitter
vacua. Finally, in the last section, we summarize our results and
mention a few interesting directions for future research.

\section{$D=5,\mathcal{N}=2$ gauged
supergravity coupled to tensor and vector multiplets}
\subsection{The ungauged theory}
The theory we consider is obtained by gauging $D=5,\mathcal{N} =
2$ supergravity coupled to vector- and tensor multiplets.  These
theories are completely determined by a constant symmetric tensor
$C_{\tI \tJ \tK}$. In particular, the manifold $\mathcal{M},$
parameterized by the $\tilde{n}$ scalars in the theory, can be
viewed as a hypersurface \be N(h)=C_{\tI \tJ \tK}h^{\tI} h^{\tJ}
h^{\tK} = 1 \ee of an ambient space with $\tilde{n}+1$ coordinates
$h^{\tI}$. The geometry of this surface is referred to as `very
special geometry'. For more details on very special geometry, see
appendix \ref{appA}.

The `very special real' manifolds were classified in \cite{DWVP}
in the case that $\mathcal{M}$ is a homogeneous space
\footnote{Homogeneous manifolds are manifolds for which its isometry group works transitively (on the manifold).  The group G of linear transformations of the $h^{\tilde{I}}$ that leave $C_{\tilde{I}\tilde{J}\tilde{K}}$ invariant is a subgroup of the isometry group $Iso(\mathcal{M})$, but it is not the whole isometry group in general.  For instance, the symmetric non-Jordan family (\ref{symnonjor}) has isometries that are not in G. Strictly speaking, only those homogeneous spaces for which G works transitively are classified. To the best of our knowledge there is no proof that there are no other homogeneous (or even symmetric) very special real spaces (for which $Iso(\mathcal{M})$ works transitively, but G not).}. The symmetric spaces are a subclass of these, and
where already found in \cite{GST1} and \cite{GST2}. They can be
divided into two subclasses, depending on whether they are
associated with Jordan algebras or not:
\begin{enumerate}
    \item When $\mathcal{M}$ is associated with a Jordan algebra, there are two subclasses:
        \begin{itemize}
            \item The `generic' or `reducible' Jordan class:
                \be \mathcal{M} = SO(1,1) \times \frac{SO(\tilde{n}-1,1)}{SO(\tilde{n}-1)},
                \qquad \tilde{n} \geq1. \ee
            \item The `irreducible' or `magical' Jordan class:
                \bea \calM &=& SL(3,\mathbb{R})/SO(3), \qquad (\tilde{n}=5) \nonumber \\
                     \calM &=& SL(3,\mathbb{C})/SU(3), \qquad (\tilde{n}=8) \nonumber \\
                     \calM &=& SU^{*}(6)/USp(6), \qquad (\tilde{n}=14) \nonumber \\
                     \calM &=& E_{6(-26)}/F_4. \qquad \qquad (\tilde{n}=26) \nonumber
                     \eea
        \end{itemize}
    \item There is one class which is not associated with a Jordan algebra, and which is therefore
    referred to as the `symmetric non-Jordan family':
        \be \calM = \frac{SO(1,\tn)}{SO(\tn)},\qquad \tn >1. \label{symnonjor}\ee
\end{enumerate}

The total global symmetry group of a matter coupled $\mathcal{N}=2$ supergravity theory factorizes into
$SU(2)_R \times G,$ where $SU(2)_R$ is the
R-symmetry group of the theory and $G$ is a group of linear transformations of the coordinates $h^{\tI}$
that leaves the tensor $C_{\tI\tJ\tK}$ invariant.
The symmetry group $G$ gives rise to isometries of the scalar manifold. In the Jordan class, $G$
even coincides with the full isometry group of $\calM$.

\subsection{Gauging the theory}

The gauging of $\mathcal{N}=2$ supergravity coupled to $n$ vector multiplets and $m$ self-dual tensor
multiplets was performed in \cite{gz99,Ceresole,5DSGREV}. The field content of the theory is
\begin{equation}
\{ e_{\mu}^{m}, \Psi_{\mu}^{i}, A_{\mu}^{I}, B_{\mu\nu}^{M},
\lambda^{i\tilde{a}}, \varphi^{\tilde{x}}\}\, ,
\end{equation}
where
\begin{eqnarray*}
I&=& 0,1, \ldots n\, ,\\
M&=& 1,2, \ldots 2m\, , \\
\tI&=& 0,1,\ldots,n+2m\, , \\
\tilde{a}&=& 1,\ldots, \tn\, ,\\
\tilde{x}&=& 1,\ldots, \tn\, ,
\end{eqnarray*}
with $\tn=n+2m$. Note that we have combined the `graviphoton' with
the $n$ vector fields of the $n$ vector multiplets into a single
$(n+1)$-plet of vector fields $A_{\mu}^{I}$ labeled by the index
$I$. Also, the spinor and scalar fields of the vector and tensor
multiplets are combined into $\tn$-tuples of spinor and scalar
fields. The indices $\tilde{a}, \tilde{b}, \ldots$ and $\tilde{x},
\tilde{y}, \ldots$ are the flat and curved indices, respectively,
of the $\tn$-dimensional target  manifold $\mathcal{M}$ of the
scalar fields. We also combine the vector and tensor indices $I$
and $M$ into one index $\tI.$

{}From the above fields, only the gravitini and the spin-1/2 fermions transform under the
$SU(2)_R$ symmetry group. However, to gauge this group we need vectors that transform in the adjoint representation of the gauge group. This problem can be solved by identifying the $SU(2)_R$ group with an $SU(2)$ subgroup of the symmetry group G of the $C_{IJK}$, and to gauge both $SU(2)$ groups simultaneously. If you just gauge a $U(1)_R$ subgroup, this problem does not occur since the adjoint of $U(1)$ is the trivial representation. An arbitrary linear combination of the vector fields can be used as $U(1)_R$ gauge field.  However, if you also gauge a subgroup K of G, the $U(1)_R$ gauge field has to be a linear combination of the K-singlet vector fields only.

The simultaneous gauging of the $U(1)_{R}$ or $SU(2)_{R}$
R-symmetry group and a subgroup $K \subset G$ introduces a scalar
potential of the form
\begin{equation}
e^{-1}\mathcal{L}_{pot}= -g^{2}P,
\end{equation}
where $P:=P^{(T)}+P^{(R)}$. $P^{(R)}$ arises from the gauging of $U(1)_{R}$ or
$SU(2)_R$, whereas $P^{(T)}$ is due to the tensor fields
transforming under the gauge group $K$ (see \bref{PLambda}).

The potential $P^{(T)}$ can be written as \cite{gzvac} \footnote{We assume $C_{MNP} = C_{IJM} = 0$. More general tensor couplings with $C_{IJM} \neq 0$ are possible, see \cite{5DSGREV}, but we will not consider these here.}
\begin{equation}
P^{(T)}=
\frac{3\sqrt{6}}{16}h^{I}\Lambda_{I}^{MN}h_{M}h_{N}\label{PLambda},
\end{equation}
with $\Lambda_{IN}^{M}$ the transformation matrices of the tensor
fields under the gauge group $K$ and \be
\Lambda_{I}^{MN}\equiv\Lambda_{IP}^{M}\Omega^{PN}=
\frac{2}{\sqrt{6}}\Omega^{MR}C_{IRP}\Omega^{PN}, \label{tenstrf}\ee with
${\stackrel{\scriptscriptstyle{\circ}}{a}}^{\tI\tJ}$ being the
inverse of ${\stackrel{\scriptscriptstyle{\circ}}{a}}_{\tI\tJ}$.
$\Omega^{MN}$ is the inverse of  $\Omega_{MN}$, which is a
(constant) invariant antisymmetric tensor of the gauge group $K$
\begin{equation}
\Omega_{MN}=-\Omega_{NM}, \qquad
\Omega_{MN}\Omega^{NP}=\delta_{M}^{P}.
\end{equation}

For the potential $P^{(R)}$ we have the following general
expression  \begin{equation}\label{PotR} P^{(R)} = -4\vec{P}\cdot
\vec{P}+ 2\vec{P}^x\cdot\vec{P}_x \, ,
\end{equation}
where $\vec{P} =h^I\vec{P}_I$ and $\vec{P}_x = h^I_{x}\vec{P}_I$ are vectors in $SU(2)$-space
(see \cite{5DSGREV}).
When we gauge the full R-symmetry group $SU(2)_{R}$ we have
\begin{equation}
\vec{P}_{I} = \vec{e}_{I}V \, ,
\end{equation}
where V is an arbitrary constant, and $\vec{e}_{I}$ are constants
that are nonzero only for I in the range of the $SU(2)$ factor and
satisfy
\begin{equation}
  \vec e_I\times \vec e_J= f_{IJ}{}^K \vec e_K \, ,
\end{equation}
with $f_{IJ}{}^K$ the $SU(2)$ structure constants. {}From now on, we will
use indices $A,B,\ldots$ for the $SU(2)$ factor. We can then take
$\vec{e}_{A} \cdot \vec{e}_{B}=\delta_{AB}$ such that (using
\bref{hhChhh}) \be P^{(R)}=-4V^2 C^{AB\tI}h_{\tI}\delta_{AB},
\label{PR=Chdelta}\ee where we defined \be
C^{\tI\tJ\tK}\equiv{\stackrel{\scriptscriptstyle{\circ}}{a}}^{\tI\tI'}
{\stackrel{\scriptscriptstyle{\circ}}{a}}^{\tJ\tJ'}
{\stackrel{\scriptscriptstyle{\circ}}{a}}^{\tK\tK'}C_{\tI'\tJ'\tK'}.
\ee

In the case of $U(1)_{R}$ gauging we have
\begin{equation}\label{U1R}
\vec{P}_{I} = V_{I}\vec{e} \, ,
\end{equation}
where $\vec{e}$ is an arbitrary vector in $SU(2)$ space and
$V_{I}$ are constants that define the linear combination of the
vector fields $A_{\mu}^{I}$ that is used as the $U(1)_R$ gauge
field
\begin{equation}\label{AU1}
A_{\mu}{[}U(1)_R{]} = V_{I}A_{\mu}^{I}.
\end{equation}
They have to be constrained by
\begin{equation}\label{Vf}
V_{I}f_{JK}^{I}=0,
\end{equation}
with $f_{JK}^{I}$ being the structure constants of $K.$ Using the
very special geometry identities of appendix \ref{appA}, the
$U(1)_{R}$ potential can be written as
\begin{equation}
P^{(R)} = -4C^{IJ\tK}V_{I}V_{J}h_{\tK} \label{P=CVVht}.
\end{equation}

Finally, we remark that when $\mathcal{M}$ is associated with a
Jordan algebra
 \cite{GST1},
one has (componentwise)
\be
C^{\tI\tJ\tK}=C_{\tI\tJ\tK}=\textrm{const.}
\ee

\section{Exploring the R-symmetry potential}
\label{app:theorem}
\subsection{U(1) R-symmetry gauging leads to tachyonic de Sitter vacua}
\label{theorem}

\subsection*{Theorem}
Without charged tensors or hypers, the potential gets only a
contribution from R-symmetry gauging.  Unlike in $4D$, non-Abelian 
vector multiplets do not contribute a term to the potential
\cite{Gunpot,GST3}. For the $U(1)_R$ case the potential is given by
\bref{P=CVVht}.

In our conventions, a critical point $\varphi_{c}$ of the
potential with $P^{(R)}(\varphi_{c})> 0$ corresponds to a de
Sitter vacuum.  We will demonstrate that if such a de Sitter
vacuum exists, it will always be unstable.  To prove this, we need
to calculate the matrix of second derivatives of the potential at
the critical point.  A critical point, by definition, obeys the
following relation \be\label{crit} \frac{\partial
P^{(R)}}{\partial \varphi^x}(\varphi_{c}) =
-4C^{IJK}V_{I}V_{J}h_{K,x}\vert_{\varphi_c} = 0\, . \ee For the mass
matrix we find \be g^{yz}\frac{\partial P^{(R)}}{\partial
\varphi^x \partial \varphi^z}\Big\vert_{\varphi_c} = \left(
\ft23P^{(R)}\delta^{y}_{x} - 8\sqrt{\ft23}V_I V_J
h^{Iu}h^{Jv}T_{uvx;z}g^{zy}\right)\Big\vert_{\varphi_c} \, , \ee
where we  have used (\ref{Txyz}), (\ref{CIJKx}) and (\ref{crit}).
We will now show that $V_{I}h^{I}_y(\varphi_c)$ is an eigenvector
of this matrix.  Equation (\ref{Txyzu}) leads to \be
T_{uvx;y}h^{Iu}h^{Jv}h^{Ky}V_{I}V_JV_K =
\sqrt{\ft32}[h^{Iu}h^{J}_{u}V_IV_JV_Kh^{K}_x -
2V_IV_Jh^{Iu}h^{Jv}T_{uv}{}^wT_{wyx}V_Kh^{Ky}]\, . \ee Since, using
(\ref{Thh}) and (\ref{crit}), we have \be
V_IV_Jh^{Iu}h^{Jv}T_{uv}{}^w \vert_{\varphi_c} = V_Ih^IV_Jh^{Jw}
\vert_{\varphi_c}\, , \ee we finally obtain \beqn
T_{uvx;y}h^{Iu}h^{Jv}h^{Ky}V_{I}V_JV_K \vert_{\varphi_c} &=&
\sqrt{\ft32}[h^{Iu}h^{J}_{u}V_IV_J -
2(V_Ih^{I})^2]V_Kh^{K}_x \vert_{\varphi_c}\, , \nonumber \\
&=&\sqrt{\ft38}P^{(R)}V_{K}h^{K}_{x} \vert_{\varphi_c}\, . \eeqn We
thus find that $V_{I}h^{I}_y(\varphi_c)$ is indeed an eigenvector
of the mass matrix with eigenvalue $-10P^{(R)}(\varphi_c)/3$.
Looking at equations (\ref{PotR}) and (\ref{U1R}) we see that if a
critical point with $P^{(R)} > 0$ exists, then also
$V_{I}h^{I}_y(\varphi_c) \neq 0$.  This proves that in case of a
de Sitter extremum, the mass matrix has always at least one
negative eigenvalue.

\subsection*{Example}
\label{example}

To illustrate our proof, we will give an example of a de Sitter
vacuum obtained by $U(1)_R$ gauging. The Jordan symmetric
spaces only lead to Anti-de Sitter or Minkowski vacua (see \cite{GST3}), but we will
show here that there are also a lot of models with de Sitter vacua.

Equations (\ref{P=CVVht}), (\ref{crit}) and (\ref{a=hh+hh}) lead to \be
\label{V=Ph} C^{IJK}V_{J}V_{K} \vert_{\varphi_c} =-\ft14
P(\varphi_c)h^{I}(\varphi_c)\, .\ee This is a necessary and
sufficient condition for $\varphi_c$ to be a critical point.
Furthermore, for $\varphi_c$ in the domain where $a_{IJ}$ is
positive definite, one can always perform a linear transformation
on the $h^I$ such that $h^{I}(\varphi_c)=(1,0,\ldots,0)$ and
$a_{IJ}(\varphi_c) =\delta_{IJ}.$ After this transformation the
polynomial $N(h)$ will take the following form \be N(h)=(h^0)^3
-\ft32 h^0 h^i h^j \delta_{ij} + C_{ijk} h^i h^j
h^k\, ,\qquad i=1,\ldots,\tn, \label{canonical}\ee which is called the canonical
parametrization of $N(h).$ Equation (\ref{V=Ph}) then becomes
\beqn C_{0JK}V_{J}V_{K}
=-\ft14 P(\varphi_c)\, , \label{C0VV=P}\\
C_{ijk}V_{j}V_{k}-V_0 V_i =0\, , \label{CijkVV-V0=0}\eeqn with
summation over repeated indices. Equation (\ref{P=CVVht}) however
leads to \be P^{(R)} (\varphi_c)=-4 C_{0JK}V_{J}V_{K} = -4V_{0}^2 + 2 \sum_{i} V_i^2 \label{C0VV=P2}\, ,\ee such
that (\ref{C0VV=P}) is automatically fulfilled. So, given a theory
(a tensor $C_{ijk}$) , we look for a vector $V_I$ that solves
equation (\ref{CijkVV-V0=0}) and for which \be
-4V_{0}^2 + 2 \sum_{i} V_i^2 <0\, .\label{cVV<0}\ee As we know, this is not
possible for general $C_{ijk}$. However, one can construct a lot
of examples. Take for example $C_{ijk}=0$.  Equation (\ref{CijkVV-V0=0}) in this case leads to $V_i = 0$, corresponding to anti-de Sitter vacua or $V_0 = 0$, corresponding to de Sitter vacua.  To study the mass matrix, we look at the particular example $n=1$. We then have
the following polynomial \be N(h)=(h^0)^3-\ft32 h^0 (h^1)^2\, .\ee
The constraint $N=1$ can be solved
by \beqn h^0 &=& \frac{\varphi}{2}+\frac{\sqrt{\varphi+\varphi^4}}{2\varphi}\\
h^1\, &=&
\sqrt{\frac{2}{3}}\left(-\frac{3}{2}\varphi+\frac{\sqrt{\varphi+\varphi^4}}{2\varphi}\right)\, .\eeqn
The metric on the scalar manifold is (using (\ref{gxy})) \be
g_{xy}=\frac{3+12\varphi^3}{4(\varphi^2+\varphi^5)}\, .\label{metricgxy}\ee
We restrict to the region $\varphi >0,$ which contains the point
$h^I(\varphi_c=1/2)=(1,0)$ and where the metric is positive
definite.

Taking $V_0=0,$ equation (\ref{CijkVV-V0=0}) is fulfilled for
arbitrary values of $V_1$ and equation (\ref{C0VV=P2}) becomes \be
P^{(R)}(\varphi_c)=2V_1^2>0\, .\label{P=2V} \ee Using the metric
(\ref{metricgxy}) and with $V_0=0$ we get for the potential \be
P^{(R)}(\varphi)=\frac{-1-8\varphi^3-40\varphi^6+12\varphi\sqrt{\varphi+\varphi^4}+24
\varphi^4\sqrt{\varphi+\varphi^4}}{2(\varphi+4\varphi^4)}V_1^2\, ,\ee
which indeed fulfils (\ref{P=2V}). Furthermore, \be
P^{(R)}_{,x}(\varphi_c)=0\, ,\ee and \be
g^{yz}P^{(R)}_{,x,z}\vert_{\varphi_c}=-\frac{20}{3}V_1^2\, ,\ee which
is indeed $-10P^{(R)}(\varphi_c)/3$ as stated in the theorem above.

\subsection{de Sitter vacua from $\mathbf{SU(2)_R}$ gauging}

For the known symmetric spaces, $SU(2)_R$ gauging never gives any critical points (see \cite{GunSU2}).
We show here that there are however also a lot of models with unstable de Sitter vacua.
Proving that there are no stable de Sitter vacua seems to be somewhat more difficult than in the $U(1)_R$ case and we hope to come back on this in a future publication.

We start from a polynomial in the canonical parametrization.
In order to gauge $SU(2)_R$, the polynomial should have an
$SU(2)_{G}$ symmetry \cite{GunSU2}. Without charged tensors this
further restricts the coefficients $C_{ijk}$ by \cite{options} \be
C_{ABC}=0\, , \qquad C_{AB\alpha}=c_{\alpha}\delta_{AB}\,  \qquad
C_{A\alpha\beta}=0\, , \label{Csu2}\ee with $c_{\alpha}$ some
arbitrary constants. We have split the indices $i=1,\ldots,n$ as  $ i=(A,\alpha)$
with $A,B,\ldots \in \{1,2,3\}$ corresponding to the SU(2) factor of the gauge group. The $C_{\alpha\beta\gamma}$ are still unconstrained.

Using expression (\ref{PR=Chdelta}) for the $SU(2)_R$ potential, the equation analogous to
(\ref{V=Ph}) is \be\label{V2=Ph}
C^{ABI}\delta_{AB}\vert_{\varphi_c} =-
P(\varphi_c)h^{I}(\varphi_c)\, .\ee We now assume that $h^{I}(\varphi_c)=(1,0,\ldots,0)$. Equation (\ref{V2=Ph}) then
leads to the conditions
\beqn
P^{(R)}(\varphi_c)=\frac{3}{2} \label{p=32}\\
c_{\alpha}=0\, ,\quad \forall\, \alpha\, , \label{cdelta=0}\eeqn and
therefore $C_{ABi}=0,\,\forall\,i$. The first condition
is again automatically fulfilled and tells us that all these critical points
are de Sitter. 

We now investigate the stability of these de Sitter vacua.
Calculating the second derivative of the potential, we get \be
P_{,x;y}=-C^{ABI}{}_{;y}\delta_{AB}h_{I,x}-C^{ABI}\delta_{AB}h_{I,x;y}\, .
\ee Using $h^{I}(\varphi_c)=(1,0,\ldots,0),$
$a_{IJ}(\varphi_c)=\delta_{IJ}$ and (\ref{hIxy}) we get for the
second term in the critical point
\be-C^{ABI}\delta_{AB}h_{I,x;y}\vert_{\varphi_c}=\frac{2}{3}P^{(R)}(\varphi_c)g_{xy}(\varphi_c)=g_{xy}(\varphi_c)\, .\ee
For the first term we also use (\ref{CIJKx}), (\ref{Txyzu}),
\bref{Txyz} and \bref{Csu2}, which gives after some calculation
\be
-C^{ABI}{}_{;y}\delta_{AB}h_{I,x}\vert_{\varphi_c}=-2g_{xy}(\varphi_c)-\ft43
h^{A}_x h^B_y \delta_{AB}\, ,\ee and therefore \be
g^{zy}P_{,x;y}\vert_{\varphi_c}=-\delta_{x}^{z}-\ft43 h^{A}_x
h^{Bz} \delta_{AB}\, .\ee This matrix has the following eigenvectors
\be \left\{ \begin{array}{l}
h^{A}_x(\varphi_c) \quad \textrm{with eigenvalue}\,\, -\ft73 \, ,\\
h^{\alpha}_x(\varphi_c) \quad \textrm{with eigenvalue}\,\, -1\, .
\end{array} \right.
\ee The de Sitter vacua are therefore always maxima of the
potential.

\section{Stable de Sitter vacua in 5D N=2 gauged super\-gravity: an
example}\label{dsexample}

The previous section made clear that $U(1)_R$ gauging alone cannot give rise to stable de Sitter vacua.  We show here that adding tensor multiplets can change this. The gauging we study in this section was already performed in \cite{gzvac} for
$\tn=3$. They found de Sitter extrema, but did not check that they
are stable. We generalize for arbitrary $\tn$ and show that the
obtained de Sitter vacua are all stable.

We consider $\mathcal{N}=2$ supergravity coupled to $\tn$ Abelian
vector multiplets and with scalar manifold
$\calM=SO(\tilde{n}-1,1) \times SO(1,1)/ SO(\tilde{n}-1),
\tilde{n} \geq1.$ The polynomial can then be written in the
following form \be N(h)=3\frac{\sqrt{3}}{2} \,
h^{0}[(h^1)^2-(h^2)^2- \ldots-(h^{\tn})^2]\, . \ee With $x=1,\ldots
,\tilde n$, introducing
\begin{equation}
  \eta _{xy}=\eta ^{xy}=\mathop{\rm Diag}\nolimits (1,-1,\ldots ,-1)\,,
 \label{defeta}
\end{equation}
we write the $C_{\tilde I\tilde J\tilde K}$ symbols as
\begin{equation}
  C_{0\underline{xy}}=\frac{\sqrt{3}}{2}\eta _{xy}\, .
 \label{Csymbols}
\end{equation}
(we underline $x$-type indices that are in fact of type $\tilde
I$, but take values in the $x$-range due to our choice). The
constraint $N=1$ can be solved by
\begin{displaymath}
h^{0}=\frac{1}{\sqrt{3}\| \varphi\|^{2}}\, , \qquad
h^{\underline{x}}=\sqrt{\frac{2}{3}}\,\varphi^{x}\, ,
\end{displaymath}
with \be\|\varphi\|^{2}=\varphi ^x\eta _{xy}\varphi ^y\, .\ee The
hypersurface $N=1$ decomposes into three disconnected
components:\\
(i) $\|\varphi\|^{2}>0$ and $\varphi^{1}>0$\\
(ii) $\|\varphi\|^{2}<0$ \\
(iii) $\|\varphi\|^{2}>0$ and $\varphi^{1}<0$.\\
In the following, we will consider the ``positive timelike''
region (i) only, since in region (ii), $g_{\tilde{x}\tilde{y}}$
and
 ${\stackrel{\scriptscriptstyle{\circ}}{a}}_{\tI\tJ}$ are not positive definite
 ,    and  region (iii) is isomorphic to region (i).

We now proceed by gauging the above theory. The isometry group of
the scalar manifold is $SO(\tilde{n}-1,1) \times SO(1,1).$ We
gauge the noncompact subgroup $SO(1,1) \subset SO(\tilde{n}-1,1)$
together with $U(1)_{R} \subset SU(2)_{R}.$ The $SO(1,1)$ subgroup
rotates $h^{1}$ and $h^{2}$ into each other and therefore acts
nontrivially on the vector fields $A_{\mu}^{1}$ and $A_{\mu}^{2}$.
In order for the resulting theory to be supersymmetric, these
vectors have to be dualized to antisymmetric tensor fields. We can
thus decompose the index $\tI$ in the following way
\be\tI=(I,M)\, ,\ee with $I,J,K,\ldots=0,3,4,\ldots,\tn$ and
$M,N,P,\ldots = 1,2.$

Furthermore, we need a vector transforming in the adjoint of $SO(1,1)$ (which
means it should be inert) to act as its gauge field. Looking at
$\Lambda_{IN}^{M} \sim C_{IRN}\Omega^{MR}$, we see that only
$A^{0}_{\mu}$ couples to the tensor fields (only $C_{0RP}\neq0$) and thus acts
as the gauge field of $SO(1,1).$ The remaining vectors are called `spectator
fields' with respect to the $SO(1,1)$ gauging.

Finally, for the $U(1)_{R}$ gauge field we take a linear
combination $A_{\mu} {[}U(1)_R{]} = V_{I}A_{\mu}^{I}$ of the
vectors. We now have the ingredients to calculate the potentials
\bref{PLambda} and \bref{P=CVVht} (taking $\Omega^{12}=
-\Omega^{21}=-1$): \bea \Lambda_{0N}^{M} &=& \frac{2}{\sqrt{6}}
\Omega^{MR}C_{0RN}= \frac{1}{\sqrt{2}} \left( \begin{array}{cc}
0 & 1\\
1 & 0
\end{array} \right)\, , \\
P^{(T)}&=&\frac{3\sqrt{6}}{16}h^{I}\Lambda_{I}^{MN}h_{M}h_{N}={1
\over
8}{(\varphi^{1})^2-(\varphi^{2})^2 \over \|\varphi\|^{6}}\, , \label{Tpotred}\\
 P^{(R)} &=&-4\sqrt{2}V_0V_i\varphi^i\|\varphi\|^{-2}+2|V|^2\|\varphi\|^2\, ,\qquad
 |V|^2\equiv V_iV_i\, , \label{Rpotred}\eea where we defined a new index $i$ as $I=(0,i).$ Then \be
P=P^{(T)}+P^{(R)}={1 \over
8}{(\varphi^{1})^2-(\varphi^{2})^2 \over \|\varphi\|^{6}}-4
\sqrt{2} V_{0}\varphi^{i}V_{i} \|\varphi\|^{-2}+2
\|\varphi\|^{2}|V|^{2}\, .\ee

Demanding $P_{,\tilde{x}}=0$ gives the following conditions on the
critical points \bea
{\varphi^{i} \over \modph^{4}}&=&16\sqrt{2} V_{0}V_{i}\, ,\label{cond1}\\
{1 \over \modph^{6}}&=&-{1\over2}\left(16\sqrt{2} V_{0}|V|
\right)^{2}+8 |V|^{2}\, , \label{cond2}\eea with the
constraints \bea
|V|^{2}\neq 0\, ,\nonumber\\
32 V_{0}^{2}<1\, .\label{constr} \eea

{}From \bref{cond2} we see that $\modph^{2}$ is completely
determined by the $V_{I}.$ {}From \bref{cond1} we see that also the
$\varphi^{i}$ are completely fixed by the $V_{I}.$ This means that
only $\varphi^{1}$ and $\varphi^{2}$ can still vary, as long as
the combination $(\varphi^{1})^{2}-(\varphi^{2})^{2}$ remains
fixed. We thus have a one parameter family of critical points,
which is due to the unbroken $SO(1,1)$. There is also an unbroken
$SO(\tilde n-3)$, but the vacuum is at the symmetric point.

The value of the potential in the critical points becomes \be
P(\varphi_c)=3 \modph^{2}|V|^{2}\left(1-32
V_{0}^{2}\right)\, , \ee which is clearly positive because of
\bref{constr} and therefore corresponds to de Sitter vacua.

We now show that all these vacua are stable. We use the index
$m=1,2$ for the scalars related to $M=1,2$. Calculating the second
derivatives of the potential in the critical points gives the
following Hessian \bea P_{,m,n}(\varphi_{c}) & = & (\eta \varphi
)_m
(\eta \varphi )_n \left[3\mf^{-8}+4\varphi^k\varphi ^k\mf^{-10}\right]\, ,\nonumber\\
P_{,m,i}(\varphi_{c})\,& = &-4(\eta
\varphi )_m \varphi ^i \left[\mf^{-8}+\varphi^k\varphi ^k\mf^{-10}\right]\, ,\nonumber\\
P_{,i,j}(\varphi_{c})\,\,\, & = & \varphi ^i\varphi
^j\left[5\mf^{-8}+4\varphi^k\varphi ^k\mf^{-10}\right]+\ft14\delta
_{ij}\mf^{-6}\, , \label{secderiv3} \eea where $(\eta \varphi
)_x\equiv \eta _{xy}\varphi ^y$.

The $SO(1,1)$ invariance implies a zero eigenvector $\varphi
^n(\sigma _1)_n{}^m$. Using this $SO(1,1)$ and the $SO(\tilde
n-2)$ of the $\varphi ^i,$ we may further use for any critical
point $\varphi_{c} = (\varphi ^1,0,\varphi ^3,0,\ldots ,0)$ with
$|\varphi ^3|<|\varphi ^1|$. Then the zero mode is $\varphi ^2$,
and the sector $\varphi ^4,\ldots ,\varphi ^{\tilde  n}$ decouples
as a unit matrix times $\ft14\mf^{-6}$. The relevant part of the
hessian therefore is
\begin{equation}
 \partial \partial P=|\mf|^{-10}
  \left(\begin{array}{cc}
 (\varphi ^1)^2\left[3(\varphi ^1)^2+(\varphi ^3)^2\right]&
  -4(\varphi ^1)^3\varphi ^3 \\ -4(\varphi ^1)^3\varphi ^3 &
  \ft14\left[(\varphi ^1)^4+18(\varphi ^1)^2(\varphi ^3)^2 -3(\varphi
  ^3)^4\right]\end{array}
   \right)\, ,
 \label{Hessmatrix}
\end{equation}
where we defined $\partial \partial P \equiv
\partial_{\tilde x}
\partial_{\tilde y} P(\varphi_c)\vert_{\tilde x,\tilde y=1,3}.$ The
determinant and trace are
\begin{equation}
  \det\partial \partial P=\ft34(\varphi ^1)^2\mf^{-14}\, ,\qquad
    \trace\partial \partial P= \ft14\mf^{-10}\left[ 13(\varphi ^1)^4
    + 22(\varphi ^1)^2(\varphi ^3)^2 - 3(\varphi ^3)^4\right]\, ,
 \label{dettrace}
\end{equation}
which shows that the eigenvalues are positive.
\subsection*{Comments}
\begin{itemize}
\item {\bf BEH effect.}  Like in \cite{stabledSa,stabledSb}, the massless scalar is a Goldstone boson.  It will get `eaten' by the SO(1,1) gauge field, making the gauge field massive.  There will thus be only positive mass scalars left in the effective theory.
\item {\bf Quantized scalar masses.}  The masses of the scalars are given by the eigenvalues of $g^{xy}P_{x,y}(\varphi_c)/P(\varphi_c)$. We already showed that these will be positive. Using Mathematica, the scalar mass spectrum turns out to be
\be
(0, \frac{8}{3}\frac{(\varphi_1)^2 - (\varphi_2)^2}{\modph^2},\frac{2}{3},\frac{2}{3},...,\frac{2}{3})\,.
\ee
In \cite{Kallosh} it was observed that all known examples of de Sitter extrema in extended supergravities have scalar masses that are quantized in units of the cosmological constant.  This is also true in our model for all scalars, but one.  One of the scalar masses depends on the parameters $V_I$ that determine the $U(1)_R$ gauge field.  Also in \cite{dsN4b} examples of (unstable) de Sitter extrema were found that had parameter-dependent scalar masses.
\item {\bf $SU(2)_R$ gauging.}  Instead of gauging $U(1)_R$ we could also have gauged the full $SU(2)_R$ R-symmetry as long as there are a sufficient number of gauge fields available ($\tilde{n} \geq 5$).  Without loss of generality, we can choose $A_{\mu}^3$, $A_{\mu}^4$, $A_{\mu}^5$ as the $SU(2)$ gauge fields.  We then find
\be
P^{(R)} = \frac{3}{2}\modph^2\,.
\ee
Looking at equation (\ref{Rpotred}), we observe that we get the same potential if we do a $U(1)_R$ gauging with $V_0=0$ and $|V|^2 = 3/4$. $SU(2)_R$ gauging with tensors charged under $SO(1,1)$ therefore will also lead to stable de Sitter vacua.  The scalar masses are in this case given by $(0,8/3,2/3, 2/3,...,2/3)$.
\end{itemize}

\section{$ U(1)_{R}$ gauging and charged tensors: more examples}
To try to find out what ingredients are really necessary to obtain
stable de Sitter vacua, we will now look at a few other examples
with $U(1)_{R}$ and charged tensors.  A natural idea is to take
the scalar manifold $\calM$ to be one of the other known symmetric
very special real manifolds.
\subsection{The magical Jordan family}
\subsubsection{$\calM = SL(3,\mathbb{R})/SO(3)$}
$\calM$ is described as the hypersurface $N(h) = 1$ of the cubic
polynomial \cite{GunSU2,DWVP} \be N(h) =
\frac{3}{2}\sqrt{3}h^3\eta_{\alpha\beta}h^{\alpha}h^{\beta} +
\frac{3\sqrt{3}}{2\sqrt{2}}\gamma_{\alpha MN}h^{\alpha}h^{M}h^{N}
\, , \ee where \beqn \alpha, \beta &=& 0,1,2 \, , \quad \qquad M,N
= 4,5 \, , \nonumber \\ \eta_{\alpha \beta} &=& diag(+,-,-)\, ,
 \quad \gamma_0 = -
\openone_2 = \left(
\begin{matrix} -1 & 0\cr
        0 & -1\end{matrix}\right)\, , \nonumber \\
 \gamma_1 &=&  \sigma_{1} = \left(
\begin{matrix} 0 & 1\cr
        1& 0\end{matrix}\right)\, , \quad \gamma_2 =\sigma_{3} =  \left(
\begin{matrix} 1 & 0\cr
        0 & -1\end{matrix}\right)\, .\eeqn
The vector field metric $a_{\tilde{I}\tilde{J}}$ becomes
degenerate when $\eta_{\alpha\beta}h^{\alpha}h^{\beta} = 0$, so we
can restrict ourselves to the region
$\eta_{\alpha\beta}h^{\alpha}h^{\beta} \neq 0$. To solve the
constraint $N(h) = 1$, we take the parametrization used in
\cite{GunSU2}, \beqn h^{\alpha} = \sqrt{\frac{2}{3}}x^{\alpha}\, ,
\, h^{M} = \sqrt{\frac{2}{3}}b^{M}\, , \, h^{3} = \frac{1-b^T
\bar{x} b}{\sqrt{3} \vert \vert x \vert\vert^2}\, , \eeqn where $b^T
\bar{x} b \equiv b^M\bar{x}_{MN}b^N$ with $\bar{x}_{MN} =
x^{\alpha}\gamma_{\alpha MN}$ and $ \vert \vert x \vert \vert^2
\equiv \eta_{\alpha \beta}x^{\alpha}x^{\beta}$.  The metrics
$a_{\tilde{I}\tilde{J}}$ and $g_{xy}$ are only positive definite
in the region $\vert \vert x \vert\vert^2 > 0$ and $x_0 > 0$.
Since this is the physically relevant region, we will restrict
ourselves to this domain from now on.

In this model we can gauge the $SO(2,1)$ symmetry between
$h^{\alpha}$ with the vector fields $A_{\mu}^{\alpha}$, while
dualizing the non-trivially charged vector fields $A_{\mu}^M$ to
tensor fields. We then get a potential \be P^{(T)} =
\frac{1}{8}b^T \bar{x}\Omega \bar{x} \Omega \bar{x}b \, , \quad
\Omega = i\sigma_{2} =
\left(\begin{matrix} 0 & 1 \cr -1 & 0 \end{matrix} \right)\, .\ee\\
Gauging the full R-symmetry is not possible in this model, but we
can gauge a $U(1)_{R}$ symmetry. We have $A_{\mu}[U(1)_R] =
V_{I}A_{\mu}^{I}$, with $V_I f_{JK}^I = 0$. {}From this it follows
that $A_{\mu}[U(1)_R] = V_{3}A_{\mu}^{3}$, so $A_{\mu}^{3}$ is the
$U(1)_R$ gauge field. Since $P^{(R)} = -4C^{IJ\tilde{K}}V_I V_J
h_{\tilde{K}}$ and $C^{33\tilde{K}} = C_{33\tilde{K}} = 0$ we find
$P^{(R)} = 0$.  The total potential $P$ is thus given by $P^{(T)}$
alone.  The critical points of $P$ are given by $b^M = 0$, leading
to Minkowski vacua.  They are supersymmetric when the $U(1)_R$
gauging is turned off ($V_3 = 0$).  There are no de Sitter vacua in this
model.

\subsubsection{$\calM = SL(3,\mathbb{C})/SU(3)$}
$\calM$ is described as the hypersurface $N(h) = 1$ of the cubic
polynomial \cite{GunSU2,DWVP} \be N(h) =
\frac{3}{2}\sqrt{3}h^4\eta_{\alpha\beta}h^{\alpha}h^{\beta} +
\frac{3\sqrt{3}}{2\sqrt{2}}\gamma_{\alpha MN}h^{\alpha}h^{M}h^{N}
\, , \ee where \beqn \alpha, \beta &=& 0,1,2,3 \, ,
\quad M,N = 5,6,7,8 \, , \nonumber \\
\eta_{\alpha \beta} &=& diag(+,-,-,-)\, , \quad \gamma_0 = - \openone_4 \, , \nonumber \\
 \gamma_1 &=&  \openone_2 \otimes \sigma_1\, , \quad \gamma_2 =  \sigma_2 \otimes \sigma_2\, , \quad \gamma_3 = \openone_2 \otimes \sigma_3\, . \eeqn
We take the same parametrization as in the previous model, \beqn
h^{\alpha} = \sqrt{\frac{2}{3}}x^{\alpha}\, , \, h^{M} =
\sqrt{\frac{2}{3}}b^{M}\, , \, h^{4} = \frac{1-b^T \bar{x}
b}{\sqrt{3} \vert \vert x \vert\vert^2}\, . \eeqn The metrics are
again only positive definite in the region $\vert \vert x
\vert\vert^2 > 0$ and $x_0 > 0$.

The model above has an $SO(3,1) \times U(1)$ symmetry, which acts on
the fields $h^{\tilde{I}}$ (and similarly on the vector fields
$A_{\mu}^{\tilde{I}}$) as \cite{DWVVP}
\begin{eqnarray}
 \delta h^\alpha  & = & B^\alpha {}_\beta h^\beta\, ,  \nonumber\\
 \delta h^M & = & \ft14 B^{\alpha \beta }(\gamma _{\alpha \beta })^M{}_N
 h^N + S^M{}_Nh^N\epsilon\, ,
 \label{deltah}
\end{eqnarray}
where
\begin{eqnarray}
  &&S\equiv \gamma _1\gamma _2\gamma _3= i\sigma _2\otimes \openone_2\, ,\qquad
S^2=-\openone_4\, ,\nonumber\\
&&  \gamma _{a b }=\gamma _{[a }\gamma _{b ]}
 =-S\varepsilon _{abc  }\gamma ^c\, ,
 \quad \gamma _{0a}=-\gamma _{a 0}=\gamma _a \, , \quad a = 1,2,3\, . 
\end{eqnarray}
Indices on the matrices $B_{\alpha \beta }$ are raised and lowered with
$\eta _{\alpha \beta }$ and these transformations satisfy $B_{\alpha
\beta }=-B_{\beta \alpha }$. This implies that they describe $\SO(1,3)$.
The motivation for the definition of $\gamma _{0a }=-\gamma _{a 0}$
is based on a larger Clifford algebra, see \cite[(5.16)]{DWVVP}.
The $\gamma $-matrices are symmetric, while $S$ is antisymmetric.
$\epsilon $ is the parameter for the $\U(1)$ symmetry.

To gauge a symmetry, we have to assign the isometry transformations to
vector multiplets, i.e. to connect the parameters of gauge transformations
$\Lambda ^I$ to parameters of the isometry group, such that the
transformations on the vector part form the adjoint representation and on
the tensor multiplets there exists an antisymmetric matrix $\Omega
_{MN}$ such that (see (\ref{tenstrf}))
\begin{equation}
  \Omega_{MP}\Lambda_{IN}^P = \frac{2}{\sqrt{6}}C_{IMN}\, .
 \label{tOmegaC}
\end{equation}
\begin{itemize}
\item
{\bf $U(1)_R$ gauging and tensors charged under $U(1)\times SU(2)$
}.

We now gauge the $\SO(3)$ part of (\ref{deltah}) with $B^{0a
}=0$ and take $A_{\mu}^{a}$ as the adjoint vectors. We gauge the $\U(1)$ by the vector field $A_{\mu}^{0}$. To gauge this symmetry, we also have to dualize the vector fields $A_{\mu}^{M}$ to tensor fields.
We have
\begin{equation}
B_{a b }= \alpha\varepsilon _{abc }\Lambda ^c\, ,\qquad \epsilon
=\beta \Lambda ^0\, , \qquad \varepsilon _{123} = 1\, ,
 \label{Beps}
\end{equation}
where we allowed for arbitrary coefficient $\alpha$ and $\beta$ to be determined below. This
leads thus to the transformation matrices
\begin{equation}
 \Lambda_{0N}^M =\beta  S^M{}_N \, , \qquad \Lambda_{a
   N}^M=-\ft12 \alpha (S\gamma _a)^M{}_N\, .
 \label{tSU2U1}
\end{equation}
Checking (\ref{tOmegaC}) gives
\begin{equation}
  \Omega =\frac{1}{\alpha} S\, ,\qquad  \beta =\frac{\alpha}{2}\, .
 \label{Omegacoeff}
\end{equation} 
For simplicity, and without losing generality, we can choose $\alpha=1$, $\beta = 1/2$.

In \cite{GunSU2} the vector fields $A_{\mu}^{a}$ where used to also gauge the full $SU(2)_R$ symmetry and it was found that the total potential has no critical points. Instead, we will use the vector field $A_{\mu}^4$ together with $A_{\mu}^0$ to gauge the $U(1)_R$
symmetry. The potential $P = P^{(T)} + P^{(R)}$ becomes \beqn
P^{(R)} &=& -2\sqrt{3}V_{0}\left ( \frac{V_0}{\sqrt{3}}\vert \vert
x \vert \vert^2 + \frac{4}{\sqrt{3}}V_4 \left(\frac{1-b^T \bar{x}
b}{\sqrt{2}\vert \vert x \vert \vert^2}\right)x^0
-\frac{2}{\sqrt{6}}V_4 b^T b \right)\,, \\
P^{(T)} &=& \frac{1}{8} b^T \bar{x}\Omega \bar{x} \Omega \bar{x} b
= - \frac{1}{8}b^T \bar{x}^3 b \nonumber \\ &=& \frac{1}{8}\vert
\vert x \vert \vert^2 b^T \bar{x}b - \frac{1}{2}x_{0}^2 b^T
\tilde{x} b + \frac{1}{4} x_{0} \vert \vert \tilde{x} \vert
\vert^2 b^T b + \frac{1}{4}x_{0}^3 b^T b\,, \quad
\label{PTsymman}\eeqn where $b^T b = b^M b^N \delta_{MN}$, $\vert
\vert \tilde{x} \vert \vert^2 = x^a x^b \delta_{ab}$, $\vert \vert
x \vert \vert^2 = \eta_{\alpha \beta}x^{\alpha}x^{\beta}$,
$\bar{x}_{MN} = x^{\alpha}\gamma_{\alpha MN}$ and  $\tilde{x}_{MN}
= x^{a}\gamma_{a MN}$.  The last line of (\ref{PTsymman}) makes
the $U(1) \times SU(2)$ symmetry of the potential manifest, since
$b^T \bar{x}b$, $b^T \tilde{x} b$, $\vert \vert \tilde{x} \vert
\vert^2$ and $\vert \vert x \vert \vert^2$ are easily seen to be
invariant under the transformations (\ref{deltah}) (with $B_{a 0} = 0$). Using this symmetry we can restrict the search for extrema to points where
e.g. $x_2 = x_3 = b_8 = 0$. We analyzed the potential with Mathematica and found no de Sitter vacua (in the region where the metrics are positive-definite).

When $b^M = 0$, finding critical points of $P$ reduces to finding critical points of $P^{(R)}$. It was shown in \cite{GST3} that $P^{(R)}$ has an Anti-de Sitter maximum if and only if $V^{\sharp \tilde{I}} \equiv \sqrt{\frac{2}{3}}C^{\tilde{I}\tilde{J}\tilde{K}}V_{\tilde{J}}V_{\tilde{K}}$ lies in the domain of positivity of the Jordan algebra, and a Minkowski critical point if and only if $V^{\sharp \tilde{I}} = 0$ ($P^{(R)}$ identically zero).  The total potential $P$ also has these critical points, but the extra potential from the tensors can change the nature of these critical points (e.g from a maximum to a saddle point).  With Mathematica we also found Anti-de Sitter vacua of $P$ with $b^M \neq 0$.  Since our primary interest was finding de Sitter vacua, we did not check the nature of these critical points.
\item {\bf
$U(1)_R$ gauging and tensors charged under $U(1)\times SO(2,1)$} \\
Instead of the compact symmetry above, we can also gauge
$U(1)\times SO(2,1)$ by again dualizing the vector fields
$A_{\mu}^{M}$ to tensor fields and choosing  the $SO(2,1)$ gauge fields to be
$A_{\mu}^{0}$, $A_{\mu}^{1}$, $A_{\mu}^{3}$, while letting
$A_{\mu}^{2}$ correspond to the $U(1)$ gauge field. Similarly as in the previous example, this leads to  $\Omega = \gamma_2 S$.  We can again
gauge the $U(1)_R$ symmetry, this time with a linear combination
of $A_{\mu}^{2}$ and $A_{\mu}^{4}$.  This leads to the following
potential, \beqn P^{(R)} &=& 2 \sqrt{3} V_2 \left
(\frac{V_2}{\sqrt{3}}\vert \vert x \vert \vert^2 -
\frac{4}{\sqrt{3}}V_4\left (\frac{1 - b^T \bar{x}b}{\sqrt{2} \vert
\vert x \vert \vert^2} \right)x^2 + \frac{2}{\sqrt{6}}V_4 b^T
\gamma_2 b \right)\,, \nonumber \\
P^{(T)} &=& \frac{1}{8} b^T \bar{x}\Omega \bar{x} \Omega \bar{x} b
\nonumber \\ &=& -\frac{1}{8}\vert \vert x \vert \vert^2 b^T
\bar{x}b - \frac{1}{2}x_{2}^2 b^T \tilde{x} b - \frac{1}{4} x_{2}
\vert \vert \tilde{x} \vert \vert^2 b^T b - \frac{1}{4}x_{2}^3 b^T
b\,, \quad \eeqn where now $\vert \vert \tilde{x} \vert
\vert^2 = (x^0)^2 - (x^1)^2 - (x^3)^2$ and $\tilde{x}_{MN} =
x^{0}\gamma_{0 MN} + x^{1}\gamma_{1 MN} + x^{3}\gamma_{3 MN}$.
Analyzing the potential as in the previous case, we again found no de Sitter vacua.  The potential has a critical point only when $V_2 = 0$. $P^{(R)}$ is then identically zero, and we have a family of Minkowski vacua at $b^M=0$.
\end{itemize}

\subsubsection{The other magical Jordan symmetric spaces}
The spaces $\calM = SU^{*}(6)/USp(6)$ and $\calM = E_{6(-26)}/F_4$
are 14 and 26 dimensional respectively, and allow for more
possibilities to get charged tensor multiplets, making the
potential more difficult to analyze.  Because all the magical
Jordan spaces have a similar structure, one might expect similar
qualitative features as in the previous models, but this has to be
checked in detail to be sure.

\subsection{The non-Jordan symmetric spaces}
We now consider theories with $\calM = {SO(1,\tn) \over SO(\tn)},
\tn >1.$ We can then take the following polynomial \be
N(h)={3\over
2}\sqrt{{3\over2}}\left(\sqrt{2}h^{0}(h^{1})^2-h^{1}\left[(h^{2})^2+\ldots+(h^{\tn})^2\right]\right)\, .\ee
This means for the non-vanishing components of the tensor
$C_{\tI\tJ\tK}$ \be C_{011}={\sqrt{3}\over 2}\, ,\quad
C_{1xy}=-{\sqrt{6}\over 4}\, \delta_{xy}\, , \,x,y = 2,\ldots,\tn\, .
\ee The constraint $N=1$ can be solved by \bea
h^{0}&=&\sqrt{{2\over
3}}\left(\frac{1}{\sqrt{2}(\varphi^1)^2}+\frac{1}{\sqrt{2}}\varphi^1
\left[(\varphi^2)^2+\ldots+(\varphi^{\tn})^2\right]\right),\\
h^{1}&=&\sqrt{\frac{2}{3}}\varphi^1, \qquad
h^{x}=\sqrt{\frac{2}{3}}\varphi^1 \varphi^x.\eea

The Lagrangian of the theory is not invariant under the full
isometry group $SO(1,\tn)$. Only the subgroup
$G=\left[SO(\tn-1)\otimes SO(1,1)\right]\ltimes T_{\tn-1}$, with
$T_{\tn-1}$ the group of translations in an $(\tn-1)$ dimensional
Euclidean space, leaves the tensor $C_{\tI\tJ\tK}$ invariant and
can thus be gauged \cite{DWVP2}. In order to gauge a subgroup
$K\subset G$ we need $Dim(K)$ vectors transforming in the adjoint
of $K.$ Furthermore, we want an additional number of vectors
transforming non-trivially under $K$. After dualization to tensor
multiplets these give the required contribution $P^{(T)}$ to the
potential.

The subgroup $SO(1,1)$ can not be gauged since all vectors
transform non-trivially under this group and we need an
invariant vector to gauge SO(1,1).

The group $SO(\tn-1)$ rotates $h^{2},\ldots,h^{\tn}$\,(and thus
also the vectors $A^{2}_{\mu},\ldots,A^{\tn}_{\mu}$) into each
other. This means that only its subgroup $SO(2)$ can be gauged in
order to have both vectors that transform in the adjoint and vectors that transform non-trivially but not in the adjoint of the gauge group. We will therefore gauge this $SO(2)$, possibly together with $SU(2)_{R}$ or $U(1)_R$. The former was already worked out in
\cite{GunSU2}, where it was shown that the potentials $P^{(R)}$
and $P=P^{(T)}+P^{(R)}$ do not have any critical points at all and
$P^{(T)}$ only has Minkowski vacua. We now investigate the latter
gauging.

We restrict ourselves to $\tn=3$, so the group $SO(2)$ acts on
$A^2_{\mu}$ and $A^3_{\mu}.$ These vectors therefore have to be
dualized to tensors. We decompose the index $\tI$ as follows
\be\tI=(I,M)\, ,\ee with $I,J,K,\ldots=0,1$ and $M,N,P,\ldots = 2,3.$
The vector $A^{1}_{\mu}$ will act as gauge field for $SO(2)$ since
its the only vector left that couples to the tensor fields. For
the $U(1)_{R}$-gauging we take the gauge vector $A_{\mu}
{[}U(1)_R{]} = V_{I}A_{\mu}^{I}.$ The constraint \bref{Vf} is
automatically fulfilled.

The potential $P^{(T)}$ then becomes (taking $\Omega^{23}=
-\Omega^{32}=-1$): \bea \Lambda_{1N}^{M} &=& {2\over \sqrt{6}}
\Omega^{MR}C_{1RN}=-{1\over 2} \left(
\begin{array}{cc}
0 & -1\\
1 & 0
\end{array} \right)\, , \\
P^{(T)}&=&\frac{3\sqrt{6}}{16}h^{I}\Lambda_{I}^{MN}h_{M}h_{N}=\frac{1}{8}(\varphi^1)^5
\left[(\varphi^2)^2+(\varphi^3)^2\right]\, .\eea The calculation of
$P^{(R)}$ however is a bit more involved since for the non-Jordan
theories $C^{\tI\tJ\tK}$ is not constant any more. The indices are
raised using $a^{\tI\tJ}$ with \be
a^{\tI\tJ}=h^{\tI}h^{\tJ}+h^{\tI}_{\tx}h^{\tJ}_{\ty}\,g^{\tx\ty}\ee
and $g^{\tx\ty}=Diag\left((\varphi^1)^2/3,\,
1/(\varphi^1)^3,\,1/(\varphi^1)^3\right).$ Then $P^{(R)}$ becomes
\be
P^{(R)}=-4C^{IJ\tK}V_{I}V_{J}h_{\tK}=-4\sqrt{2}\frac{V_{0}V_{1}}{\varphi^1}
-(\varphi^1)^2\left[V_{0}\left((\varphi^2)^2+(\varphi^3)^2
\right)+\sqrt{2}V_1\right]^2.\ee We remark that we have to
restrict to $\varphi^1 > 0$ in order for $g_{\tx\ty}$ to be positive
definite.

As already mentioned, $P^{(T)}$ only has Minkowski ground states. Moreover, $P^{(R)}$ can at most have unstable de Sitter vacua, as we proved in section \ref{theorem}. We will now study the total potential $P^{(T)} + P^{(R)}$.

\subsubsection*{The critical points of $P$}
The total potential is \be P=\frac{1}{8}(\varphi^1)^5
\left[(\varphi^2)^2+(\varphi^3)^2\right]-4\sqrt{2}\frac{V_{0}V_{1}}{\varphi^1}
-(\varphi^1)^2 A^2\, ,\ee with \be
A=V_0 \left[(\varphi^2)^2+(\varphi^3)^2 \right]+\sqrt{2} V_1\, . \ee
The first derivatives are \bea
P_{,1}&=&\ft{5}{8}(\varphi^1)^4
\left[(\varphi^2)^2+(\varphi^3)^2\right]+4\sqrt{2}\frac{V_0
V_1}{(\varphi^1)^2}-2\varphi^1 A^2\, , \label{ptot1}\\
P_{,2}&=& \ft14 (\varphi^1)^2 \varphi^2 \left[(\varphi^1)^3-16 A V_0\right]\, ,  \label{ptot2}\\
P_{,3}&=& \ft14 (\varphi^1)^2 \varphi^3 \left[(\varphi^1)^3-16 A
V_0\right]\, . \label{ptot3}\eea {}From \bref{ptot2} and \bref{ptot3}
we get the following three possibilities for the critical points:
\begin{itemize}
\item When $\varphi^2=\varphi^3=0$ and $V_1 = 0$, equations \bref{ptot1}-\bref{ptot3} are fulfilled and  $P(\varphi_c) = 0$, giving Minkowski vacua.  Since $V_Ih^I_{\tilde{x}}(\varphi_c) \neq 0$, supersymmetry is broken unless also $V_0 = 0$.  
\item When $\varphi^2=\varphi^3=0$ and $V_1 \neq 0$, equation\bref{ptot1} leads to the condition 
\be (\varphi^1)^3=\sqrt{2}
\frac{V_0}{V_1}\, . \label{phi13V0V1}\ee
Then \be
P(\varphi_c)=-6 (\varphi^1)^2 V_{1}^2 \label{PRphic}\ee and
we have an Anti-de Sitter vacuum. The vectors
$V_{I}h^{I}_{\tx}(\varphi_c)$ and $h_{Mx}\Omega^{MN}h_N(\varphi_c)$ are now identically zero, which means the vacuum preserves the full $N=2$ supersymmetry. \item The third possibility is
$(\varphi^1)^3=16 A V_0,$ which can be rewritten as \be
V_1=\frac{p-16 V_0^2 q}{16 \sqrt{2} V_0}\, , \label{V1=frac}\ee with
$p=(\varphi^1)^3$ and $q=(\varphi^2)^2+(\varphi^3)^2 > 0.$ Using this
in (\ref{ptot1}) and solving for $V_0$ gives us the following four
solutions \be V_0=\pm\frac{1}{8} \sqrt{5p^2+\frac{2p}{q}\pm
\frac{\sqrt{4p^2+12p^3q+25p^4q^2}}{q}}\, .\label{V0=sqrt}\ee Remark
that the expressions under the square roots are always positive. We substitute (\ref{V1=frac}) and (\ref{V0=sqrt}) into
the potential $P$ and get \be P(\varphi_c)=
\frac{3p^{5/3}q(p+5p^2q\pm\sqrt{4p^2+12p^3q+25p^4q^2})}{4\left(2p+5p^2q
\pm \sqrt{4p^2+12p^3q+25p^4q^2}\right)}\, .\ee Here both signs are
positive when the second sign choice in (\ref{V0=sqrt}) is
positive, otherwise both signs are negative (independent of the
choice of the first sign in (\ref{V0=sqrt})). With the plus signs
we have a de Sitter vacuum, with the minus signs an anti-de Sitter
vacuum.

Calculating the matrix of second derivatives $P,x,y$ and
substituting (\ref{V1=frac}) and (\ref{V0=sqrt}), we get the
following form \be P,x,y(\varphi_c)=\left(\begin{array}{cc}
B_{2\times2}&0\\
0&0
\end{array}\right)\, .\ee The expected zero eigenvalue from the $SO(1,1)$ invariance
is already explicit. Furthermore, \be Det(B_{2\times2})=
-\ft{3}{32}p^{5/3} q \left(14p+25p^2
q\pm5\sqrt{4p^2+12p^3q+25p^4q^2}\right)\, ,\ee where again the plus
sign corresponds to the de Sitter vacuum, the minus sign to the
anti-de Sitter vacuum. The determinant is always negative, so
there is always a negative eigenvalue and the de Sitter vacua are unstable.

\end{itemize}

\subsection{Conclusions}
These examples seem to suggest that the existence of stable de
Sitter vacua is very model dependent.  A $U(1)_R$ gauging and
tensors charged under a non-compact gauge group do not guarantee
stable de Sitter vacua.  On the other hand, we also found a de
Sitter vacuum in a model with $U(1)_R$ gauging and tensors charged
under a compact group.  Unfortunately the de Sitter vacuum was
unstable.  Whether this is a general feature of compact gaugings
is not clear to us.

\section{Stable de Sitter vacua with hypers}
Our goal in this section is to show that it is still possible to
get stable de Sitter vacua when hypermultiplets are included. We
will do this by giving a particular example, namely we will gauge
a specific isometry of the universal hypermultiplet.  There are
probably many other possibilities, but we will not analyse this in
its generality here.

When there are charged hypers in the model, the potential gets
some extra contributions.  The total potential is given by
\cite{Ceresole,5DSGREV}\be P = -4\vec{P}\cdot \vec{P}+
2\vec{P}^x\cdot\vec{P}_x +2\mathcal{N}_{iA}\mathcal{N}^{iA} +
P^{(T)}\, , \label{Pothyp}\ee where, as before, $\vec{P} =h^I\vec{P}_I$,
$\vec{P}_x = h^I_{x}\vec{P}_I$ and ${\mathcal N}^{iA} =
\ft{\sqrt6}{4} h^I K_I^X f_X^{iA}$. Here $f_X^{iA}$ are the
quaternionic vielbeins, $f_X^{iA}f_{YiA} = g_{XY}$ with $g_{XY}$
the metric of the quaternionic-K\" ahler hypermultiplet scalar
manifold, $K_I^{X}$ are the Killing vectors and $\vec{P}_I$ the
corresponding prepotentials.

The metric of the universal hypermultiplet, together with the
Killing vectors and corresponding prepotentials were given in
\cite{universal}, and we will repeat the results here for
convenience of the reader. The four hyperscalars $q^{X}$ are
labelled as $\{V,\sigma,\theta,\tau\}$ and the metric is given by
\begin{equation}
\rmd  s^2 = \frac{\rmd V^2}{2V^2} + \frac{1}{2V^2}\left( \rmd \sigma + 2
\theta \, \rmd  \tau - 2 \tau \, \rmd  \theta\right)^2 + \frac{2}{V} \,
\left(\rmd \tau^2 + \rmd  \theta^2\right) \,. \label{quatmetric}
\end{equation}
The determinant for this metric is $1/V^6$ and since the metric is
positive definite in $\theta = \tau = 0$ if $V>0$, the metric will
be positive-definite and well-behaved everywhere as long as $V>0$.
This parametrization of the universal hypermultiplet is the one
that comes out naturally from the Calabi-Yau compactifications of
M-theory, where V acquires the meaning of the volume of the
Calabi-Yau manifold.  The metric (\ref{quatmetric}) has an SU(2,1)
isometry group generated by the following eight Killing vectors
$k_{\alpha}^X$
\begin{equation}
\begin{array}{l}
\vec{k}_1 = \left(
\begin{array}{c}
0 \\ 1 \\ 0 \\ 0
\end{array}
\right)\,,
\quad
\vec{k}_2 = \left(
\begin{array}{c}
0 \\ 2\theta \\ 0 \\ 1
\end{array}
\right)\,,
\quad
\vec{k}_3 = \left(
\begin{array}{c}
0 \\ -2\tau \\ 1 \\ 0
\end{array}
\right)\,,
\quad
\vec{k}_4 = \left(
\begin{array}{c}
0 \\ 0 \\ -\tau \\ \theta
\end{array}
\right)\,, \\
\vec{k}_{5} = \left(
\begin{array}{c}
V \\ \sigma \\ \theta/2 \\ \tau/2
\end{array}
\right)\,,
\quad
\vec{k}_{6} =\left(
\begin{array}{c}
2V \sigma  \\ \sigma^2 - \left(V + \theta^2 + \tau^2\right)^2  \\
\sigma \theta - \tau \left(V + \theta^2 + \tau^2\right)\\
\sigma \tau + \theta \left(V + \theta^2 + \tau^2\right)
\end{array}
\right)\,,
\\
\vec{k}_{7} =\left(
\begin{array}{c}
- 2V \theta\\ -\sigma\theta + V\tau + \tau\left(\theta^2 + \tau^2\right) \\
\frac{1}{2} \left(V  - \theta^2 + 3\tau^2\right) \\
-2 \theta \tau - \sigma/2
\end{array}
\right)\,,
\quad
\vec{k}_{8} =\left(
\begin{array}{c}
- 2V \tau\\ -\sigma\tau - V\theta - \theta\left(\theta^2 + \tau^2\right)  \\
-2 \theta \tau + \sigma/2\\
\frac{1}{2}\left(  V  + 3\theta^2 - \tau^2\right)
\end{array}
\right)\, .
\end{array}
\label{killvec}
\end{equation}
The corresponding prepotentials $P^{r}_{\alpha}$ are given by
\begin{equation}
\begin{array}{rcl}
\vec{P}_{1} = \left(
\begin{array}{c}  0 \\ 0  \\ -\frac{1}{4V} \end{array}
\right)\,,
\quad
\vec{P}_{2} = \left(
\begin{array}{c} -\frac{1}{\sqrt{V}} \\ 0  \\ -\frac{\theta}{V} \end{array}
\right)\,,
\quad
\vec{P}_{3} = \left(
\begin{array}{c}  0 \\ \frac{1}{\sqrt{V}}  \\ \frac{\tau}{V}\end{array}
\right)\,,
\quad
\vec{P}_{4} = \left(
\begin{array}{c}  -\frac{\theta}{\sqrt{V}} \\ -\frac{\tau}{\sqrt{V}} \\
\frac{1}{2} - \frac{\theta^2+ \tau^2}{2V} \end{array}
\right)\,, \\
\vec{P}_{5} = \left(
\begin{array}{c}  -\frac{\tau}{2\sqrt{V}} \\ \frac{\theta}{2\sqrt{V}}
\\ -\frac{\sigma}{4V} \end{array}
\right)\,,
\quad
\vec{P}_{6} = \left(
\begin{array}{c} -\frac{1}{\sqrt{V}}\left[\sigma\tau +
\theta\left(-V+\theta^2+ \tau^2\right)\right] \\
\frac{1}{\sqrt{V}}\left[\sigma\theta -
\tau\left(-V+\theta^2+ \tau^2\right)\right]    \\
-\frac{V}{4}-\frac{1}{4V}\left[\sigma^2 + \left(\theta^2+
\tau^2\right)^2\right] + \frac{3}{2}\left(\theta^2+ \tau^2\right)
\end{array} \right)\,,
\\
\vec{P}_{7} = \left(
\begin{array}{c}  \frac{4\theta\tau + \sigma}{2\sqrt{V}} \\
\frac{3\tau^2- \theta^2}{2\sqrt{V}} - \frac{\sqrt{V}}{2}  \\
-\frac{3}{2}\tau+\frac{1}{2V}\left[\sigma\theta+ \tau \left(\theta^2+
\tau^2\right)\right]\end{array} \right)\,, \quad \vec{P}_{8} = \left(
\begin{array}{c} -\frac{3 \theta^2-\tau^2}{2\sqrt{V}} + \frac{\sqrt{V}}{2}  \\
     \frac{\sigma-4\theta\tau }{2\sqrt{V}} \\
\frac{3}{2}\theta+\frac{1}{2V}\left[\sigma\tau- \theta \left(\theta^2+
\tau^2\right)\right] \end{array} \right)\,.
\end{array}
\label{prep}
\end{equation}
The Killing vectors $K_I^X$ are now given by
$V_{I}^{\alpha}k_{\alpha}^X$, where the components of the embedding matrix $V_{I}^{\alpha}$ are constants
that determine which isometries are gauged and which vector fields
are used to gauge them. The corresponding prepotentials
$\vec{P}_{I}$ then become $V_{I}^{\alpha}\vec{P}_{\alpha}$.

We are now ready to give a concrete example.  We choose to gauge
the $U(1)$ (hypermultiplet) isometry generated by $2\vec{k}_4 -
\vec{k}_1 - \vec{k}_6$, so we take \be \vec{K}_I = V_I (2\vec{k}_4
-\vec{k}_1 - \vec{k}_6) \, ,\quad \vec{P}_I = V_I (2\vec{P}_4
-\vec{P}_1 - \vec{P}_6) = V_I \vec{Q}\, , \ee where we have defined $\vec{Q} \equiv 2\vec{P}_4
-\vec{P}_1 - \vec{P}_6$. For the scalar manifold of the vector multiplets we choose $\calM=SO(\tilde{n}-1,1) \times SO(1,1)/ SO(\tilde{n}-1), \tilde{n} \geq1$, and again gauge a noncompact $SO(1,1)$ isometry of this manifold by dualizing the two charged
vector fields to tensor fields (see section \ref{dsexample} for
notation and more details).  For simplicity, we do not charge the
hypers under this symmetry. Our gauge group is thus $SO(1,1)\times
U(1)$, where two tensors are charged under the $SO(1,1)$  and the
hypers are charged under the $U(1)$. We then find that
\beqn
\frac{\partial P}{\partial q^X}(\varphi, q_c)= 0 \, , \label{crithyp} \\
P(\varphi, q_{c}) = \frac{9}{4}P^{(R)}(\varphi) + P^{(T)}(\varphi)\, , \label{Pothypcrit} \eeqn where $q_{c} =
\{V=1,\sigma =0,\theta = 0, \tau = 0\}$ and with $P^{(R)}$ and
$P^{(T)}$ given in equations (\ref{Rpotred}) and (\ref{Tpotred})
respectively. 

To verify this, first notice that $K_I^X\vert_{q_c} = 0$ and therefore 
\be
\mathcal{N}_{iA}\vert_{q_c} = 0 \, , \qquad \frac{\partial (\mathcal{N}_{iA}\mathcal{N}^{iA})}{\partial q^X}\vert_{q_c} = 0\, .
\ee
We also have $\frac{\partial P^{(T)}}{\partial q^X} = 0$ since $P^{(T)}$ only depends on the scalars of the vector multiplets.  The remaining term $-4\vec{P}\cdot \vec{P}+
2\vec{P}^x\cdot\vec{P}_x$ in equation (\ref{Pothyp}) can be written as
\be
-4\vec{P}\cdot \vec{P}(\varphi,q) + 2\vec{P}^x\cdot\vec{P}_x(\varphi,q) = -4C^{IJK}V_IV_Jh_K(\varphi)\vec{Q}\cdot\vec{Q}(q)\, ,
\ee
which shows that the $\varphi$ (vector multiplet) and $q$ (hypermultiplet) dependence of this part factorizes.
Since $\vec{Q}\vert_{q_c} = (0,0,3/2)$, to verify equation (\ref{crithyp}) we only need to check that $\frac{\partial Q^3}{\partial q^X}\vert_{q_c} = 0$, with $Q^3$ the third component of the vector $\vec{Q}$. Because $Q^3$ is quadratic in $\theta$, $\sigma$ and $\tau$ we have \be\frac{\partial Q^3}{\partial \theta} = \frac{\partial Q^3}{\partial \sigma} = \frac{\partial Q^3}{\partial \tau} = 0\quad  \textrm{if} \quad \theta = \sigma = \tau = 0\, . \ee Finally \be Q^3\vert_{\theta = \sigma = \tau = 0} = 1 + 1/4V + V/4 \, ,\ee and it's easy to see that $V=1$ is an extremum. This proofs equations (\ref{crithyp}) en (\ref{Pothypcrit}).
 
We thus find that in the point $q_{c}$, up to a
factor $9/4$ which can be absorbed in the $V_{I}$, the
potential reduces to the same potential for the vector multiplet
scalars as found in section \ref{dsexample}, where we gauged a
$U(1)_{R}$ symmetry. We now have to calculate the mass matrix in
the critical point.  Since (\ref{crithyp}) is true for any value
of the vector multiplet scalars, we have \be \frac{\partial^2
P}{\partial q^X
\partial \varphi^x}\vert_{q_{c}} = 0 \, , \ee so we can calculate the
masses of the hyper-scalars and the vector-scalars separately.
Since we already calculated the mass matrix for the
vector multiplet scalars, we just have to calculate the mass matrix
for the hypers. After a straightforward calculation we find the
matrix to be diagonal, with all entries always positive. There are
only 2 different diagonal elements and they can both be written as
a sum of manifestly positive terms.  Because the expressions are
quite messy, we will not give them in their generality here. Instead
we will look at the particular case $V_{0} = 0$.  We then have
$\varphi_{i} = 0$ in the critical point and the expressions
simplify significantly.  Concretely, we find \beqn
\frac{\partial_X\partial^Y P}{P}\Big\vert_{c} = \left(
\begin{matrix} 8/9 & 0 & 0 & 0\cr
        0 & 8/9 & 0 & 0 \cr
        0 & 0 & 4/9 & 0 \cr 0 & 0 & 0 & 4/9\end{matrix}\right) \, ,
\eeqn where derivation with respect to $q^X$ is denoted by
$\partial_X$ and indices are raised with the (inverse)
quaternionic-K\" ahler metric $g^{XY}$.  This shows that
potentials with stable de Sitter vacua also exist when hypers are
included.

\section{Summary}

In this paper we investigated different possibilities to get
stable de Sitter vacua in $5D$ $N=2$ gauged supergravity. We
proved that $U(1)_R$ gauging (without tensors) at most leads to
unstable de Sitter vacua. In the case of $SU(2)_R$ gauging we found lots
of theories exhibiting de Sitter extrema, but were unable to find
stable de Sitter vacua. However, by also introducing tensor multiplets
and gauging a non-compact symmetry group together with the
R-symmetry group we managed to construct 5D supergravity theories
with stable de Sitter vacua. The used ingredients are however not
sufficient to guarantee stable de Sitter vacua, as the analysis of
several other examples made clear. Finally, we showed with a
specific example that we can also get stable de Sitter vacua if we
replace the R-symmetry gauging with charged hypers.

There are several directions in which we plan to continue our research. First of all it would be interesting to find out under which conditions stable de Sitter vacua exist in supergravity theories. A more general analysis of the potentials coming from $SU(2)_R$
gauging and tensors will certainly be useful for this.
Investigating the potential coming from charged hypermultiplets
might also give interesting results.  Another possible fruitful
path would be to try to embed the stable de Sitter vacua in $N=4$ and
$N=8$ supergravity and check whether they are still stable. Attempts in this direction in 4D have failed (see \cite{dsN4b}), and it would be interesting to see whether this generalizes to higher dimensions.  Having an N=8 embedding could also make it easier to find their 10 or 11 dimensional origin, if any.  Finally, considering the similarities between 4D
and 5D $N=2$ supergravity, the results we found perhaps suggest
that investigating 4D gauged supergravities with tensor couplings
might lead to new examples of stable de Sitter vacua. We hope to come
back on these issues in the near future. \\

\noindent{ \bf Acknowledgements}\vspace{0.3cm}

We would like to thank A. Van Proeyen for helpful discussions and for proofreading this manuscript.  We have also greatly benefited from conversations with A. Celi and want to thank M. Trigiante for useful e-mail correspondence. This work is supported in part by the European Community's Human Potential Programme under contract MRTN-CT-2004-005104 `Constituents,
fundamental forces and symmetries of the universe', by the FWO - Vlaanderen, project
G.0235.05 and by the Federal Office for Scientific, Technical and Cultural Affairs through the `Interuniversity Attraction Poles Programme -- Belgian Science Policy' P5/27.

\appendix
\section{Very special real geometry} \label{appA}
In this appendix we will repeat some elements of very special real
geometry for convenience of the reader. This presentation is
mostly based on the appendix in \cite{5DSGREV} and the classic
paper on $5D$ $N=2$ supergravity \cite{GST1}.
\label{app:veryspecial}

Very special real manifolds are the scalar manifolds of $N=2$
$D=5$ supergravity coupled to vector(/tensor) multiplets and are
completely determined by a symmetric $3$-tensor $C_{IJK}$. Let M
be the following $n+1$ dimensional subspace of $\mathbb{R}^{n+1}$
\be M = \{h^{I} \in \mathbb{R}^{n+1} \vert N(h) =
C_{IJK}h^{I}h^{J}h^{K} > 0 \}\, , \ee with metric \be a_{IJ} = -
\frac{1}{3}
\partial_{I}\partial_{J} \ln{N(h)}\, . \ee Then the very special real
manifold $\mathcal{M}_{n}$ can be defined as the hypersurface
$N(h) = 1$ with metric the induced metric from the embedding space
M, \be g_{xy} =
\frac{3}{2}{\stackrel{\scriptscriptstyle{\circ}}{a}}_{IJ}h^{I}_{,x}h^{J}_{,y}
= -3C_{IJK}h^I h^{J}_{,x}h^{K}_{,y}\, , \label{gxy}\ee with
$h^{I}(\phi)$ obeying $C_{IJK}h^{I}(\phi)h^{J}(\phi)h^{K}(\phi) =
1$, $,x$ denoting an ordinary derivative with respect to $\phi^x$
and \footnote{It should be understood that the $h^I$ obey $N(h) =
1$ from here on.} \be
{\stackrel{\scriptscriptstyle{\circ}}{a}}_{IJ} \equiv
a_{IJ}\vert_{N=1} = -2C_{IJK}h^K + 3h_I h_J\, , \quad  h_I \equiv
C_{IJK}h^J h^K = {\stackrel{\scriptscriptstyle{\circ}}{a}}_{IJ}h^J\, .
\label{defquantI} \ee Defining
\begin{equation}
  h^I_x\equiv -\sqrt{\ft32}h^I_{,x}(\phi )\,,
 \label{hIx}
\end{equation}
we have $h_Ih^I_x=0$, leading to
\begin{equation}
  h_{Ix}\equiv {\stackrel{\scriptscriptstyle{\circ}}{a}}_{IJ}h^J_x=\sqrt{\ft32} h_{I,x}(\phi )\, , \quad  h^I h_{Ix} = 0\, .
 \label{lowerhIx}
\end{equation}
Using the above relations we can also write
${\stackrel{\scriptscriptstyle{\circ}}{a}}_{IJ}$ as \be
{\stackrel{\scriptscriptstyle{\circ}}{a}}_{IJ} = h_I h_J + h_I^x
h_{Jx}\, , \label{a=hh+hh} \ee and we have
\begin{eqnarray}
 h_{Ix;y} & = & \sqrt{\ft23}\left( h_I g_{xy}+T_{xyz}h^z_I\right)\, ,  \nonumber\\
 h^I{}_{x;y} & = & -\sqrt{\ft23}\left( h^I
 g_{xy}+T_{xyz}h^{Iz}\right)\, , \label{hIxy}\end{eqnarray}
where `;' is a covariant derivative using the Christoffel
connection calculated from the metric $g_{xy}$, with
 \be
 T_{xyz}  \equiv  C_{IJK}h^I_xh^J_yh_z^K\, .
 \label{Txyz}
\ee The previous relations can be used to derive some useful
identities.

Comparing (\ref{a=hh+hh}) and (\ref{defquantI}), we obtain
\begin{equation}
  h_I^xh_{Jx}=-2C_{IJK}h^K+2h_Ih_J\,,
 \label{hhChhh}
\end{equation}
and taking the  covariant derivative with respect to $\phi ^y$ of
(\ref{hhChhh}) leads to
\begin{equation}
  T_{xyz}h_I^xh_J^z=C_{IJL}h^L_y+h_{(I}h_{J)y}\,.
 \label{Thh}
\end{equation}
Finally, after a straightforward calculation, we get \beqn
T_{xyz;u}& = & \sqrt{\ft32}[g_{(xy}g_{z)u}-2T_{(xy}{}^wT_{z)uw}] \, , \label{Txyzu}\\
 C^{IJK}{}_{,x}& = & 2T_{uvw;x}h^{Iu}h^{Jv}h^{Kw} \,.
 \label{CIJKx}
\eeqn This formula was found in \cite{GST3}, but with an erroneous
factor $3$ instead of $2$.

The domain of the variables should be limited to $h^I(\phi )\neq
0$ and the metrics
${\stackrel{\scriptscriptstyle{\circ}}{a}}_{IJ}$ and $g_{xy}$
should be positive definite. The latter two conditions are
equivalent.

\begingroup\raggedright

\endgroup

\end{document}